\documentclass[12pt]{article}
\usepackage{graphicx}

\def\hybrid{\topmargin 0pt      \oddsidemargin 0pt
        \headheight 0pt \headsep 0pt
       \voffset-1cm
        \textwidth 6.25in       
       \textheight 9.5in       
        \marginparwidth 0.0in
        \parskip 5pt plus 1pt   \jot = 1.5ex}
\catcode`\@=11
\def\marginnote#1{}

\newcount\hour
\newcount\minute
\newtoks\amorpm
\hour=\time\divide\hour by60
\minute=\time{\multiply\hour by60 \global\advance\minute by-\hour}
\edef\standardtime{{\ifnum\hour<12 \global\amorpm={am}%
        \else\global\amorpm={pm}\advance\hour by-12 \fi
        \ifnum\hour=0 \hour=12 \fi
        \number\hour:\ifnum\minute<10 0\fi\number\minute\the\amorpm}}
\edef\militarytime{\number\hour:\ifnum\minute<10 0\fi\number\minute}

\def\draftlabel#1{{\@bsphack\if@filesw {\let\thepage\relax
   \xdef\@gtempa{\write\@auxout{\string
      \newlabel{#1}{{\@currentlabel}{\thepage}}}}}\@gtempa
   \if@nobreak \ifvmode\nobreak\fi\fi\fi\@esphack}
        \gdef\@eqnlabel{#1}}
\def\@eqnlabel{}
\def\@vacuum{}
\def\draftmarginnote#1{\marginpar{\raggedright\scriptsize\tt#1}}

\def\draftlabel#1{{\@bsphack\if@filesw {\let\thepage\relax
   \xdef\@gtempa{\write\@auxout{\string
      \newlabel{#1}{{\@currentlabel}{\thepage}}}}}\@gtempa
   \if@nobreak \ifvmode\nobreak\fi\fi\fi\@esphack}
        \gdef\@eqnlabel{#1}}
\def\@eqnlabel{}
\def\@vacuum{}
\def\draftmarginnote#1{\marginpar{\raggedright\scriptsize\tt#1}}

\def\draft{\oddsidemargin -.5truein
        \def\@oddfoot{\sl preliminary draft \hfil
        \rm\thepage\hfil\sl\today\quad\militarytime}
        \let\@evenfoot\@oddfoot \overfullrule 3pt
        \let\label=\draftlabel
        \let\marginnote=\draftmarginnote
   \def\@eqnnum{(\theequation)\rlap{\kern\marginparsep\tt\@eqnlabel}%
\global\let\@eqnlabel\@vacuum}  }

\def\numberbysection{\@addtoreset{equation}{section}
        \def\theequation{\thesection.\arabic{equation}}}

\def\underline#1{\relax\ifmmode\@@underline#1\else
        $\@@underline{\hbox{#1}}$\relax\fi}

\def\titlepage{\@restonecolfalse\if@twocolumn\@restonecoltrue\onecolumn
     \else \newpage \fi \thispagestyle{empty}\c@page\z@
        \def\thefootnote{\fnsymbol{footnote}} }

\def\endtitlepage{\if@restonecol\twocolumn \else  \fi
        \def\thefootnote{\arabic{footnote}}
        \setcounter{footnote}{0}}  
\relax

\hybrid

\newfont{\Bbb}{msbm10 scaled 1\@ptsize00}
\newfont{\Bbbb}{msbm7 scaled 1\@ptsize00}

\newcommand{\DDD}{\raise-1pt\hbox{$\mbox{\Bbbb D}$}}

\newcommand{\UUU}{\raise-1pt\hbox{$\mbox{\Bbbb U}$}}

\newcommand{\ZZ}{\mbox{\Bbb Z}}
\newcommand{\z}{\raise-1pt\hbox{$\mbox{\Bbbb Z}$}}

\newcommand{\sss}{\raise-1pt\hbox{$\mbox{\Bbbb S}$}}
\newcommand{\ccc}{\raise-1pt\hbox{$\mbox{\Bbbb C}$}}

\def\beq{\begin{equation}}
\def\eeq{\end{equation}}
\def\p{\partial}
\def\todd{{\bf t}_{o}}
\def\te{{\bf t}_{e}}
\def\toprime{{\bf t}_{o}'}

\def\T{{\bf T}}

\newtheorem{theorem}{Theorem}
\newtheorem{lemma}{Lemma}
\newtheorem{lemma-definition}{Lemma-Definition}

\newtheorem{remark}{Remark}

\begin{document}

\begin{titlepage}

\title{Toda chain solves large BKP}

\author{
A.~Zabrodin\thanks{
National Research University Higher School of Economics,
20 Myasnitskaya Ulitsa,
Moscow 101000, Russia; 
e-mail: zabrodin@itep.ru}}

\date{October 2026}
\maketitle


\begin{abstract}

We prove that any solution of the Toda chain hierarchy (the 
1D reduction of the 2D Toda lattice) is a solution of the 
large BKP hierarchy. This is a difference
analogue of Alexandrov's result that KdV solves (small) BKP.

\end{abstract}

\end{titlepage}

\vspace{5mm}

%

\tableofcontents

\vspace{5mm}

\section{Introduction}

In the seminal paper \cite{A21}, 
Alexandrov obtained an interesting and rather unexpected result 
that any solution to the KdV hierarchy is also a solution to the 
BKP hierarchy (after a simple rescaling of variables).
The aim of the present note is to extend this result by establishing 
a similar correspondence between difference analogues of the two
hierarchies which are the Toda chain and large BKP (the latter
is also known as $B$-Toda lattice).

The Toda chain \cite{Toda1,Toda2} is one of the most famous 
integrable systems from which their modern theory originated.
It is of great interest from various points of view,
both mathematical and physical. Equations of motion for this system are 
second order differential equations for a variable
$\varphi (n, t)$ which depends on a lattice point $n\in \ZZ$ and 
time $t$, and can be interpreted either as a coordinate of 
$n$-th particle on the line or a ``field'' in a $(1+1)$-dimensional 
``space-time'', where the space coordinate $n$ is discrete.
The equations of motion are differential in time 
$t$ and difference in $n$. 
It is well known that a properly taken continuous limit 
converts the Toda chain into the (even more famous) 
KdV equation for a function $u(x,t)$, with the space variable
$x$ arising from the lattice variable $n$ in the limit.
Moreover, this can be extended to the whole
infinite hierarchies of the KdV and Toda chain equations.
In our present work, we treat the Toda chain as a 1D reduction
of the 2D Toda lattice \cite{UT84}, just as KdV is a
reduction of KP.

The hierarchy which we call ``large BKP'' was first mentioned
in \cite{JM83} but was not further elaborated there. 
It was independently introduced and studied 
in \cite{KL98} under the name
``charged BKP''. Later, the authors of \cite{OST12,OST16}
(see also \cite{vLO15,book})
suggested to call it ``large BKP'' as opposed to 
the small (or ``neutral'') BKP which
was introduced in 
\cite{JM83,DJKM83} and further studied in 
\cite{DJKM82}-\cite{Z21}. 
It was this ``small'' BKP hierarchy that 
Alexandrov dealt with in his work \cite{A21}.
Recently, the large BKP was rediscovered in 
\cite{KZ22,PZ23} in an essentially different context. Namely,
it was constructed as a subhierarchy of
the 2D Toda lattice and called there the Toda lattice of type B (B-Toda).
As such, it can be regarded as a natural integrable discretization of the
small BKP hierarchy, just as the 2D Toda lattice is a (partial) 
discretization of KP. 

We will work in 
the bilinear formalism (originally developed 
by Hirota \cite{Hirotabook})
in the most general form that was given to it in the 
works of the Kyoto school \cite{JM83,DJKM83}. In this approach,
the main object of interest is {\it tau-function} which 
plays the role of a universal dependent variable meaning that
all dependent variables (such as $\varphi (n,t)$ for the
Toda chain or $u(x,t)$ for KdV) can be expressed through it.
The tau-function is a function
of infinitely many variables, which are parameters of commuting
flows of the hierarchy (``times''). Our main technical tools will be
one or another version of the integral bilinear relation for the
tau-function. This relation 
encodes all equations of the hierarchy.

Let us briefly recall Alexandrov's result.
To do this, let’s start with the KP hierarchy in the bilinear 
formalism \cite{JM83}. The set of independent variables (times) is
${\bf t}=\{t_1, t_2, t_3, \ldots \}$. The tau-function
$\tau^{\rm KP}({\bf t})$ satisfies the bilinear integral
relation
\beq\label{kp1}
\oint_{C_{\infty}}\! dz\,
e^{\xi ({\bf t} -{\bf t}' ,z)}
\tau^{\rm KP} \Bigl ({\bf t} -[z^{-1}] \Bigr )
\tau^{\rm KP} \Bigl ({\bf t}' +[z^{-1}]\Bigr )=0,
\eeq
valid for all ${\bf t}$, ${\bf t}'$.
Here $C_{\infty}$ is a big circle of large enough radius $R$
(a ``small'' contour around infinity) and the following 
standard notations are used:
\beq\label{not}
\begin{array}{l}
\displaystyle{
\xi ({\bf t}, z)=\sum_{k\geq 1} t_k z^k,}
\\ \\
{\bf t}\pm [z^{-1}]=\Bigl \{t_1 \pm z^{-1}, t_2 \pm \frac{1}{2}z^{-2},
t_3 \pm \frac{1}{3}z^{-3}, \, \ldots \}.
\end{array}
\eeq

The reduction in which the tau-function does not depend on ``even'' times,
i.e., 
\beq\label{red}
\frac{\p   \tau^{\rm KP}({\bf t})}{\p \, t_{2k}}\,=0 \;\;
\mbox{for all ${\bf t}$ and all $k\geq 1$},
\eeq
is called the KdV hierarchy.
The independent variables are
$
\todd =\{t_1, t_3, t_5, \ldots \}
$
(the ``odd'' times). The tau-function $\tau^{\rm KdV}$ is defined as
\beq\label{taukdv}
\tau^{\rm KdV}(\todd )=\tau^{\rm KP}({\bf t}),
\eeq
where the ``even'' times $\te =\{t_2, t_4, \ldots \}$ 
in the right-hand side may take any
values since the reduction just means that 
$\tau^{\rm KP}({\bf t})$ does not depend on them.

The set of independent variables of the small 
BKP hierarchy is just the set of ``odd'' times
$
\todd
$, the same as in the KdV case.
The tau-function of the small BKP hierarchy,
$\tau^{\rm b} (\todd )$, 
satisfies the integral bilinear equation of the form
\beq\label{BB1}
\frac{1}{2\pi i}\oint_{C_{\infty}}\, \frac{dz}{z}\,
e^{\xi_{o} (\todd -\toprime ,z)}
\tau^{\rm b} \Bigl (\todd -2[z^{-1}]_{o}\Bigr )
\tau^{\rm b} \Bigl (\toprime +2[z^{-1}]_{o}\Bigr )=
\tau^{\rm b} (\todd )\tau^{\rm b} (\toprime ),
\eeq
valid for all $\todd , \, \toprime $. Here we use the 
following notation (the ``odd'' version of (\ref{not})):
\beq\label{not1}
\begin{array}{l}
\displaystyle{
\xi _{o} (\todd ,z)=\sum_{k \geq 1, \, {\rm odd}}
t_k z^k,}
\\ \\ 
\todd \pm 2[z^{-1}]_{o} =\Bigl \{t_1 \pm 2z^{-1}, \,
t_3 \pm \frac{2}{3} z^{-3}, \, t_5 \pm \frac{2}{5} z^{-5},
\ldots \, \Bigr \}.
\end{array}
\eeq

A simple and elegant argument from Alexandrov’s paper \cite{A21}
allows one to prove the following statement connecting solutions 
of the two hierarchies.

\begin{theorem}(\cite{A21})
\label{theorem:main0}
For any KdV tau-function $\tau^{\rm KdV}(\todd )$ 
the function
\beq\label{int0}
\tau^{\rm b}(\todd )=\tau^{\rm KdV}(\todd /2 )
\eeq
is a tau-function of the small BKP hierarchy.
\end{theorem}

As mentioned above,
the purpose of the present note is to generalize this result
to the ``difference analogues'' of the KdV and (small) BKP -- 
to the Toda chain (TC) and the large BKP hierarchies. 
The independent variables in this case are
$\T =\{ T_1, T_2, T_3, \ldots \}$ (an infinite set of continuous 
times) and a discrete variable $n\in \ZZ$. The corresponding 
tau-functions
are $\hat \tau^{\rm TC}(n, \T )$ and $\hat \tau^{\rm B}(n, \T )$
respectively. The analogue of Theorem \ref{theorem:main0} 
proved in this note is as
follows:

\begin{theorem}
\label{theorem:main1}
For any TC tau-function $\hat \tau^{\rm TC}(s, \T )$ 
the function
\beq\label{int1}
\hat \tau^{\rm B}(n, \T )=\left \{
\begin{array}{l}
\hat \tau^{\rm TC}(\frac{n}{2}, \frac{\T}{2} ), \quad n\in 2\ZZ ,
\\ \\
\hat \tau^{\rm TC}( \frac{n+1}{2}, \frac{\T}{2}), \quad n\in 2\ZZ +1
\end{array} \right.
\eeq
is a tau-function of the large BKP hierarchy.
\end{theorem}

\noindent
It is worth emphasizing that the properly taken continuous limit
in $n$ reproduces Alexandrov's result (\ref{int0}).

\begin{remark}
Perhaps it is helpful to represent
(\ref{int1}) in a more readable form:

{\footnotesize
$$
\begin{array}{ccccccc}
\cdots  \hspace{-0.2cm}& \hat \tau^{\rm B}(2k\! -\! 3, \T )=
\hat \tau^{\rm B}(2k \! -\! 2, \T )\hspace{-0.2cm} &  \mbox{---}
 & \hspace{-0.2cm}
\hat \tau^{\rm B}(2k\! -\! 1, \T )=
\hat \tau^{\rm B}(2k, \T ) \hspace{-0.2cm} &  \mbox{---} 
& \hspace{-0.2cm} 
\hat \tau^{\rm B}(2k\! +\! 1, \T )=
\hat \tau^{\rm B}(2k \! +\! 2, \T ) & \hspace{-0.2cm} \cdots
\\ &&&&&& \\
& \| & & \| & & \| &
\\ &&&&&& \\
& \hat \tau^{\rm TC}(k\! -\! 1, \T /2 ) & &
\hat \tau^{\rm TC}(k, \T /2 ) & &
\hat \tau^{\rm TC}(k\! +\! 1, \T /2 ) &
\end{array}
$$
}

\noindent
So, we see that solutions to the large BKP 
delivered by the Toda chain are rather
special: the values of $\hat \tau^{\rm B}$
at neighboring lattice points $2k-1$ and $2k$ are the same for 
any $k\in \ZZ$ (but in general depend on $k$).
\end{remark}

Throughout the paper, we will deal with several different
tau-functions. For reader's convenience, we list them here:
\begin{itemize}
\item[] $\tau^{\rm KP}({\bf t})$ -- tau-function of the KP hierarchy,
\item[] $\tau^{\rm KdV}(\todd )$ -- tau-function of the KdV hierarchy (reduction of KP),
\item[] $\tau^{\rm TL}(s, {\bf t}, \bar {\bf t})$ -- 
tau-function of the Toda lattice (TL) hierarchy,
\item[] $\tau^{\rm TC}(s, {\bf T} )$ -- tau-function of the 
Toda chain hierarchy (reduction of TL),
\item[] $\tau^{\rm b}(\todd )$ -- tau-function of the 
small BKP hierarchy,
\item[] $\tau^{\rm B}(n, {\bf t})$ -- tau-function of the 
large BKP hierarchy.
\end{itemize}
Interrelations between them can be illustrated by the 
following diagram:

{\large
\beq\label{diagram}
\begin{array}{ccc}
\tau^{\rm KP} & \stackrel{{\rm difference}}{\longrightarrow} 
& \tau^{\rm TL}
\\ 
\downarrow\lefteqn{{\scriptstyle \rm red}} & & 
\downarrow\lefteqn{{\scriptstyle \rm red}} 
\\ 
\tau^{\rm KdV} & \stackrel{{\rm difference}}{\longrightarrow} 
& \tau^{\rm TC}
\\
\uparrow  & &  \uparrow
\\ 
\tau^{\rm b} & \stackrel{{\rm difference}}{\longrightarrow} 
& \tau^{\rm B}
\end{array}
\eeq
}

\noindent
The horizontal arrows mean integrable discretization, t.e., passing
to discrete (or difference) analogues of the 
hierarchies in the left column. The downward-pointing arrows mean
reductions. The upward-pointing arrows in the bottom row symbolize the fact that any solution to hierarchies in the middle row 
is a solution to hierarchies in the bottom row. In particular, 
the upward-pointing arrow in the right column symbolizes the 
main result of the present note.

In addition, it should be borne in mind that there 
are other relationships between the hierarchies 
discussed that are not shown in the diagram. Namely, the small BKP
is a subhierarchy of KP and large BKP is a subhierarchy of 
the TL.

Lastly, we note that instead of the tau-functions 
$\tau^{\rm TL}$, $\tau^{\rm TC}$, $\tau^{\rm B}$ from the right
column it will be more convenient to deal with their slightly
modified versions
$\hat \tau^{\rm TL}$, $\hat \tau^{\rm TC}$, $\hat \tau^{\rm B}$
which differ by a simple times-dependent factor (see (\ref{TL2}),
(\ref{L2}) below).

\section{Toda lattice and Toda chain}

The TL  hierarchy \cite{UT84}
plays a very important role in the theory of integrable
systems. The commuting flows of the hierarchy 
are parameterized by two infinite sets of complex times
${\bf t}=\{t_1, t_2, t_3, \ldots \}$ 
(``positive times'') and $\bar {\bf t}=\{\bar t_1,
\bar t_2, \bar t_3, \ldots \}$ (``negative times''), 
together with the integer-valued ``zeroth time'' $s\in \ZZ$. 
(Hereafter, the bar does not
mean complex conjugation.) 
Equations of the hierarchy are differential in the times 
${\bf t}$, $\bar {\bf t}$ and
difference in $s$. 

The universal dependent variable is the tau-function
$\tau^{\rm TL}(s, {\bf t}, \bar {\bf t})$. All equations 
of the hierarchy are encoded in the bilinear integral
relation
\beq\label{TL1}
\begin{array}{l}
\displaystyle{
\oint_{C_{\infty}}\frac{dz}{z}\, z^{s-s'} 
e^{\xi ({\bf t}-{\bf t}', z)}
\tau^{\rm TL}(s, {\bf t}-[z^{-1}], \bar {\bf t})
\tau^{\rm TL}(s'+1, {\bf t}'+[z^{-1}], \bar {\bf t}')}
\\ \\
\phantom{aaaaaaaaa}
\displaystyle{=
\oint_{C_{\infty}}\frac{dz}{z}\, z^{s'-s} 
e^{\xi (\bar {\bf t}-\bar {\bf t}', z)}
\tau^{\rm TL}(s+1, {\bf t}, \bar {\bf t}-[z^{-1}])
\tau^{\rm TL}(s', {\bf t}', \bar {\bf t}'+[z^{-1}])}
\end{array}
\eeq
valid for all ${\bf t}$, ${\bf t}'$ and $s, s' \in \ZZ$.

For our purpose it will be more convenient to deal with
a slightly modified tau-function:
\beq\label{TL2}
\hat \tau^{\rm TL}(s, {\bf t}, \bar {\bf t})=
\exp \Bigl (\sum_{k\geq 1}kt_k \bar t_k \Bigr )\,
\tau^{\rm TL}(s, {\bf t}, \bar {\bf t}).
\eeq
The bilinear relation for the $\hat \tau^{\rm TL}$ has the form
\beq\label{TL3}
\begin{array}{l}
\displaystyle{
\oint_{C_{\infty}}\frac{dz}{z}\, z^{s-s'} 
e^{\xi ({\bf t}-{\bf t}', z)+\xi (\bar {\bf t}-\bar {\bf t}', z^{-1})}
\hat \tau^{\rm TL}(s, {\bf t}-[z^{-1}], \bar {\bf t})
\hat \tau^{\rm TL}(s'+1, {\bf t}'+[z^{-1}], \bar {\bf t}')}
\\ \\
\phantom{aaaa}
\displaystyle{=
\oint_{C_{\infty}}\frac{dz}{z}\, z^{s'-s} 
e^{\xi (\bar {\bf t}-\bar {\bf t}', z)+\xi ( {\bf t}-{\bf t}', z^{-1})}
\hat \tau^{\rm TL}(s+1, {\bf t}, \bar {\bf t}-[z^{-1}])
\hat \tau^{\rm TL}(s', {\bf t}', \bar {\bf t}'+[z^{-1}])}.
\end{array}
\eeq

\begin{remark}
In terms of the tau-function $\hat \tau^{\rm TL}$, the simplest
nonzero 
solution to the hierarchy is 
$\hat \tau^{\rm TL}(s, {\bf t}, \bar {\bf t})=1$.
\end{remark}

Passing to the 1D Toda chain, we introduce the new time
variables ${\bf T}=\{ T_1, T_2, \ldots \}$ and 
${\bf Y}=\{ Y_1, Y_2, \ldots \}$ that are linear combinations
of ${\bf t}$ and $\bar {\bf t}$:
\beq\label{TL4}
\left \{ \begin{array}{l}
{\bf T}=\frac{1}{2}({\bf t}-\bar {\bf t}),
\\ \\
{\bf Y}=\frac{1}{2}({\bf t}+\bar {\bf t}),
\end{array}\right.
\quad 
\mbox{or} \;\;\;
\left \{ \begin{array}{l}
{\bf t}={\bf Y}+{\bf T},
\\ \\
\bar {\bf t}={\bf Y}-{\bf T}.
\end{array}\right.
\eeq
The reduction to the 1D TC means that the tau-function
$\hat \tau^{\rm TL}(s, {\bf Y}+{\bf T},  {\bf Y}-{\bf T})$
does not depend
on the ${\bf Y}$-variables, and thus depends on 
the ${\bf T}$-variables only (and on $s$). 
This allows us to introduce the
tau-function of the TC hierarchy 
$\hat \tau^{\rm TC}(s, {\bf T})$ as follows:
\beq\label{TL5}
\hat \tau^{\rm TC}(s, {\bf T})=
\hat \tau^{\rm TL}(s, {\bf Y}+{\bf T},  {\bf Y}-{\bf T}).
\eeq
(The ${\bf Y}$-variables here may take any values since the reduction
just means that the tau-function does not depend on them.)
In terms of this tau-function, the bilinear relation (\ref{TL3})
for the TC hierarchy acquires the form
\beq\label{TL6}
\begin{array}{l}
\displaystyle{
\oint_{C_{\infty}}\!\! \frac{dz}{z} z^{s-s'} 
e^{\xi ({\bf T}-{\bf T}', z)-\xi ({\bf T}-
{\bf T}', z^{-1})+g(z)}
\hat \tau^{\rm TC}\Bigl (s, {\bf T}-{\textstyle \frac{1}{2}}[z^{-1}]
\Bigr )
\hat \tau^{\rm TC}\Bigl (s'+1, {\bf T}'+{\textstyle \frac{1}{2}}[z^{-1}]
\Bigr )}
\\ \\
\displaystyle{=
\oint_{C_{\infty}}\!\! \frac{dz}{z} z^{s'-s} 
e^{\xi ( {\bf T}-{\bf T}', z^{-1})-\xi ( {\bf T}-{\bf T}', z)+g(z)}
\hat \tau^{\rm TC}\Bigl (s\! +\! 1, {\bf T}\! +\! 
{\textstyle \frac{1}{2}} [z^{-1}]\Bigr )
\hat \tau^{\rm TC}\Bigl (s', {\bf T}' \! -\! {\textstyle \frac{1}{2}}[z^{-1}]\Bigr )},
\end{array}
\eeq
where
$
g(z)=\xi ({\bf Y}-{\bf Y}', z)+\xi ({\bf Y}-{\bf Y}', z^{-1})
$. Since the ${\bf Y}$-variables are arbitrary, $g(z)$ 
in (\ref{TL6}) is
an arbitrary function of the form 
$$
g(z)=\sum_{k\geq 0}a_k(z^k + z^{-k})=g(z^{-1}),
$$
and equation (\ref{TL6}) holds for any such function.
Equivalently, this arbitrary function can be represented
as a series
\beq\label{arb}
g(z)=\sum_{k\geq 0} b_k (z+z^{-1})^k,
\eeq
with some other coefficients $b_k$.

\section{Large BKP}
\label{section:dLargeBKP}

The set of independent variables of the large BKP hierarchy 
is the infinite set of continuous
times
${\bf t}=\{t_1, t_2, t_3, \ldots \}$ and the discrete
variable $n\in \ZZ$. 
The bilinear equation for the tau-function $\tau^{\rm B} (n, {\bf t})$
has the form
\beq\label{L1}
\begin{array}{ll}
&\displaystyle{
\frac{1}{2\pi i} \oint_{C_{\infty}} \frac{dz}{z^2}\, 
z^{n-n'} e^{\xi ({\bf t}-{\bf t}', z)}
\tau^{\rm B} \Bigl ( n-1, {\bf t}-[z^{-1}]\Bigr )
\tau^{\rm B} \Bigl ( n'+1, {\bf t}'+[z^{-1}]\Bigr )}
\\ &\\
+& \displaystyle{
\frac{1}{2\pi i} \oint_{C_{\infty}} \frac{dz}{z^2}\, 
z^{n'-n} e^{-\xi ({\bf t}-{\bf t}', z)}
\tau^{\rm B} \Bigl ( n+1, {\bf t}+[z^{-1}]\Bigr )
\tau^{\rm B} \Bigl ( n'-1, {\bf t}'-[z^{-1}]\Bigr )}
\\ &\\
=& \frac{1}{2} \Bigl (1-(-1)^{n-n'}\Bigr )
\tau^{\rm B} (n, {\bf t})\tau^{\rm B} (n', {\bf t}'),
\end{array}
\eeq
which is valid for all $n, n'$, ${\bf t}, {\bf t}'$.

\begin{remark}\label{remark:dkp}
If $n-n'\in 2\ZZ$, the right-hand side of (\ref{L1}) vanishes.
In this case the bilinear relation splits into two
disconnected ones, each of them 
connecting tau-functions with $n$ of the same parity. This
defines the DKP hierarchy (known
also as Pfaff lattice or coupled KP \cite{HO}-\cite{Takasaki07}), 
which can be regarded as a subhierarchy
of the large BKP one. 
\end{remark}

\begin{lemma}
The function
$$
\tau^{\rm B} \Bigl ( n, {\bf t}\Bigr )=
\exp \Bigl (\frac{1}{2}\sum_{k\geq 1}k t_k^2\Bigr )
$$
is a solution to equation (\ref{L1}).
\end{lemma}

\noindent
For the proof, see \cite{PZ23}. This is the simplest possible
solution, which
suggests introducing a slightly modified tau function:
\beq\label{L2}
\hat \tau^{\rm B} \Bigl ( n, {\bf t}\Bigr )=
\exp \Bigl (-\frac{1}{2}\sum_{k\geq 1}k t_k^2\Bigr )
\tau^{\rm B} \Bigl ( n, {\bf t}\Bigr ),
\eeq
for which the simplest solution is 
$\hat \tau^{\rm B} \Bigl ( n, {\bf t}\Bigr )=1$.
The bilinear relation for it can be straightforwardly obtained
from (\ref{L1}). It has the form
\beq\label{L1a}
\begin{array}{ll}
&\displaystyle{
\frac{1}{2\pi i} \oint_{C_{\infty}} \! \frac{dz}{z^2-1}\, 
z^{n-n'} e^{\xi ({\bf t}-{\bf t}', z)
-\xi ({\bf t}-{\bf t}', z^{-1})}
\hat \tau^{\rm B} \Bigl ( n-1, {\bf t}-[z^{-1}]\Bigr )
\hat \tau^{\rm B} \Bigl ( n'+1, {\bf t}'+[z^{-1}]\Bigr )}
\\ &\\
+& \displaystyle{
\frac{1}{2\pi i} \oint_{C_{\infty}} \! \frac{dz}{z^2-1}\, 
z^{n'-n} e^{-\xi ({\bf t}-{\bf t}', z)+
\xi ({\bf t}-{\bf t}', z^{-1})}
\hat \tau^{\rm B} \Bigl ( n+1, {\bf t}+[z^{-1}]\Bigr )
\hat \tau^{\rm B} \Bigl ( n'-1, {\bf t}'-[z^{-1}]\Bigr )}
\\ &\\
=& \frac{1}{2} \Bigl (1-(-1)^{n-n'}\Bigr )
\hat \tau^{\rm B} (n, {\bf t})\hat \tau^{\rm B} (n', {\bf t}').
\end{array}
\eeq

\section{Proof of the main theorem}

Now we are ready to prove the main theorem 
(Theorem \ref{theorem:main1}).
We should prove that any solution $\hat \tau^{\rm TC}(s, \T )$ to 
equation (\ref{TL6}) solves equation (\ref{L1a}) after 
the identification of variables as in (\ref{int1}).

We start with the case when both $n$ and $n'$ in (\ref{L1a})
are odd, so its right-hand side vanishes. After the 
simultaneous shifts
$n \to n+1$, $n'\to n'+1$ the equation acquires the form
\beq\label{L1b}
\begin{array}{l}
\displaystyle{
\oint_{C_{\infty}} \frac{dz}{z^2-1}\, 
z^{n-n'} e^{\xi ({\bf t}-{\bf t}', z)-\xi ({\bf t}-{\bf t}', z^{-1})}
\hat \tau^{\rm B} \Bigl ( n, {\bf t}-[z^{-1}]\Bigr )
\hat \tau^{\rm B} \Bigl ( n'+2, {\bf t}'+[z^{-1}]\Bigr )}
\\ \\
+\, \displaystyle{
\oint_{C_{\infty}} \frac{dz}{z^2-1}\, 
z^{n'-n} e^{-\xi ({\bf t}-{\bf t}', z)+\xi ({\bf t}-{\bf t}', z^{-1})}
\hat \tau^{\rm B} \Bigl ( n+2, {\bf t}+[z^{-1}]\Bigr )
\hat \tau^{\rm B} \Bigl ( n', {\bf t}'-[z^{-1}]\Bigr )}=0,
\end{array}
\eeq
which is valid for all ${\bf t}, {\bf t}'$ and $n, n' \in 2\ZZ$.
Our goal is to identify it with (\ref{TL6}). To this end, we put
$s=n/2$, $s'=n'/2$, $\T ={\bf t}/2$, $\T ' ={\bf t}'/2$
and rewrite (\ref{TL6}) in the form
\beq\label{L2a}
\begin{array}{l}
\displaystyle{
\oint_{C_{\infty}}\!\! \frac{dz}{z} \, z^{\frac{1}{2}(n-n')} 
e^{\frac{1}{2}\xi ({\bf t}-{\bf t}', z)-\frac{1}{2} \xi ({\bf t}-
{\bf t}', z^{-1})+g(z)}\,
\hat \tau^{\rm TC}\Bigl ({\textstyle \frac{n}{2}}, 
{\textstyle \frac{{\bf t}-[z^{-1}]}{2}}
\Bigr )\,
\hat \tau^{\rm TC}\Bigl ({\textstyle \frac{n'+2}{2}}, 
{\textstyle \frac{{\bf t}'+[z^{-1}]}{2}}
\Bigr )}
\\ \\
\displaystyle{=
\oint_{C_{\infty}}\!\! \frac{dz}{z} \, z^{\frac{1}{2}(n'-n)} 
e^{\frac{1}{2}\xi ( {\bf t}-{\bf T}', z^{-1})-
\frac{1}{2}\xi ( {\bf t}-{\bf t}', z)+g(z)}\,
\hat \tau^{\rm TC}\Bigl ({\textstyle \frac{n+2}{2}}, 
{\textstyle \frac{{\bf t} + 
[z^{-1}]}{2}}\Bigr )\,
\hat \tau^{\rm TC}\Bigl ({\textstyle \frac{n'}{2}},  
{\textstyle \frac{{\bf t}'  - [z^{-1}]}{2}}\Bigr )}.
\end{array}
\eeq
The main step is to choose the arbitrary function $g(z)$ properly.
The proper choice is
\beq\label{g1}
e^{g(z)}=\frac{\varphi (z) -\varphi (z^{-1})}{z-z^{-1}},
\eeq
where
\beq\label{g2}
\varphi (z)=z^{\frac{1}{2}(n-n')} 
e^{\frac{1}{2}\xi ({\bf t}-{\bf t}', z)-\frac{1}{2} 
\xi ({\bf t}- {\bf t}', z^{-1})}.
\eeq
It is clear that the function $g(z)$ chosen in this way has the 
acceptable general form
(\ref{arb}). Plugging it into (\ref{L2a}), we have:
\beq\label{L3}
\begin{array}{l}
\displaystyle{
\oint_{C_{\infty}}\! \frac{dz}{z^2-1} \, z^{n-n'} 
e^{\xi ({\bf t}-{\bf t}', z)-\xi ({\bf t}-
{\bf t}', z^{-1})}\,
\hat \tau^{\rm TC}\Bigl ({\textstyle \frac{n}{2}}, 
{\textstyle \frac{{\bf t}-[z^{-1}]}{2}}
\Bigr )\,
\hat \tau^{\rm TC}\Bigl ({\textstyle \frac{n'+2}{2}}, 
{\textstyle \frac{{\bf t}'+[z^{-1}]}{2}}
\Bigr )}
\\ \\
\phantom{aaaaaaaaa}\displaystyle{
-\, \oint_{C_{\infty}}\! \frac{dz}{z^2-1} \, 
\hat \tau^{\rm TC}\Bigl ({\textstyle \frac{n}{2}}, 
{\textstyle \frac{{\bf t}-[z^{-1}]}{2}}
\Bigr )\,
\hat \tau^{\rm TC}\Bigl ({\textstyle \frac{n'+2}{2}}, 
{\textstyle \frac{{\bf t}'+[z^{-1}]}{2}}
\Bigr )}
\\ \\
\displaystyle{=-\, 
\oint_{C_{\infty}}\! \frac{dz}{z^2-1} \, z^{n'-n} 
e^{-\xi ({\bf t}-{\bf T}', z)+
\xi ( {\bf t}-{\bf t}', z^{-1})}\,
\hat \tau^{\rm TC}\Bigl ({\textstyle \frac{n+2}{2}}, 
{\textstyle \frac{{\bf t} + 
[z^{-1}]}{2}}\Bigr )\,
\hat \tau^{\rm TC}\Bigl ({\textstyle \frac{n'}{2}},  
{\textstyle \frac{{\bf t}'  - [z^{-1}]}{2}}\Bigr )}
\\ \\
\phantom{aaaaaaaaa}\displaystyle{+\, 
\oint_{C_{\infty}}\! \frac{dz}{z^2-1} \,
\hat \tau^{\rm TC}\Bigl ({\textstyle \frac{n+2}{2}}, 
{\textstyle \frac{{\bf t} + 
[z^{-1}]}{2}}\Bigr )\,
\hat \tau^{\rm TC}\Bigl ({\textstyle \frac{n'}{2}},  
{\textstyle \frac{{\bf t}'  - [z^{-1}]}{2}}\Bigr )}.
\end{array}
\eeq
The second integrals in both sides vanish because the residue
at infinity is zero. So, we are left with
\beq\label{L3a}
\begin{array}{l}
\displaystyle{
\oint_{C_{\infty}}\! \frac{dz}{z^2-1} \, z^{n-n'} 
e^{\xi ({\bf t}-{\bf t}', z)-\xi ({\bf t}-
{\bf t}', z^{-1})}\,
\hat \tau^{\rm TC}\Bigl ({\textstyle \frac{n}{2}}, 
{\textstyle \frac{{\bf t}-[z^{-1}]}{2}}
\Bigr )\,
\hat \tau^{\rm TC}\Bigl ({\textstyle \frac{n'+2}{2}}, 
{\textstyle \frac{{\bf t}'+[z^{-1}]}{2}}
\Bigr )}
\\ \\
\displaystyle{+\,
\oint_{C_{\infty}}\! \frac{dz}{z^2-1} \, z^{n'-n} 
e^{-\xi ({\bf t}-{\bf T}', z)+
\xi ( {\bf t}-{\bf t}', z^{-1})}\,
\hat \tau^{\rm TC}\Bigl ({\textstyle \frac{n+2}{2}}, 
{\textstyle \frac{{\bf t} + 
[z^{-1}]}{2}}\Bigr )\,
\hat \tau^{\rm TC}\Bigl ({\textstyle \frac{n'}{2}},  
{\textstyle \frac{{\bf t}'  - [z^{-1}]}{2}}\Bigr )}=0.
\end{array}
\eeq
Now it is seen that after setting
\beq
\hat \tau^{\rm TC}\Bigl (n/2, 
{\bf t}/2\Bigr )=
\hat \tau^{\rm B}(n, {\bf t}), \quad n\in 2\ZZ ,
\eeq
equation (\ref{L3a}) coincides with (\ref{L1b}).
This gives a solution to (\ref{L1b}) for even values of $n$ 
(the even sublattice), i.e., a solution to the DKP hierarchy.

To extend this solution to odd values of $n$,
we set
\beq\label{ans}
\hat \tau^{\rm B}(n-1, {\bf t})=
\hat \tau^{\rm B}(n, {\bf t}), \quad n\in 2\ZZ .
\eeq
Then, setting 
$s=\frac{1}{2}(n+1)$, $s'= \frac{1}{2}(n'+1)$, 
$\T ={\bf t}/2$, $\T ' ={\bf t}'/2$ in (\ref{TL6}), 
we can repeat the above
calculation almost literally and conclude that 
$$
\begin{array}{l}
\hat \tau^{\rm TC} \Bigl (\frac{n+1}{2}, \frac{{\bf t}}{2}\Bigr )
=\hat \tau^{\rm B} (n, {\bf t}), \quad n\in 2\ZZ +1,
\end{array}
$$
solves equation (\ref{L1b}).

Passing to other possible cases, 
let us rewrite equation (\ref{L1a}) for the case when $n$ and
$n'$ have different parities:
\beq\label{L1c}
\begin{array}{l}
\displaystyle{
\phantom{aa}\oint_{C_{\infty}} \frac{dz}{z^2-1}\, 
z^{n-n'} e^{\xi ({\bf t}-{\bf t}', z)-\xi ({\bf t}-{\bf t}', z^{-1})}
\hat \tau^{\rm B} \Bigl ( n-1, {\bf t}-[z^{-1}]\Bigr )
\hat \tau^{\rm B} \Bigl ( n'+1, {\bf t}'+[z^{-1}]\Bigr )}
\\ \\
\phantom{a}+\, \displaystyle{
\oint_{C_{\infty}} \frac{dz}{z^2-1}\, 
z^{n'-n} e^{-\xi ({\bf t}-{\bf t}', z)+\xi ({\bf t}-{\bf t}', z^{-1})}
\hat \tau^{\rm B} \Bigl ( n+1, {\bf t}+[z^{-1}]\Bigr )
\hat \tau^{\rm B} \Bigl ( n'-1, {\bf t}'-[z^{-1}]\Bigr )}
\\ \\
=\, \tau^{\rm B}(n, {\bf t})\tau^{\rm B}(n', {\bf t}').
\end{array}
\eeq
Again, our aim is to identify it with (\ref{TL6}).
Choosing
\beq\label{g1a}
e^{g(z)}=\frac{\psi (z) -\psi (z^{-1})}{z-z^{-1}},
\eeq
where
\beq\label{g2a}
\psi (z)=z^{\frac{1}{2}(s-s'-1)} 
e^{\frac{1}{2}\xi ({\bf t}-{\bf t}', z)-\frac{1}{2} 
\xi ({\bf t}- {\bf t}', z^{-1})},
\eeq
we can see that (\ref{TL6}) converts into
\beq\label{L10}
\begin{array}{l}
\displaystyle{
\oint_{C_{\infty}} \frac{dz}{z^2 -1} z^{2(s-s')-1}\,
e^{\xi ({\bf t}-{\bf t}', z)-\xi ({\bf t}-
{\bf t}', z^{-1})}\,
\hat \tau^{\rm TC}\Bigl (s, 
{\textstyle \frac{{\bf t}-[z^{-1}]}{2}}
\Bigr )\,
\hat \tau^{\rm TC}\Bigl (s'+1, 
{\textstyle \frac{{\bf t}'+[z^{-1}]}{2}}
\Bigr )}
\\ \\
+\, \displaystyle{
\oint_{C_{\infty}} \frac{dz}{z^2 -1} z^{-2(s-s')+1}\,
e^{-\xi ({\bf t}-{\bf t}', z)-\xi ({\bf t}-
{\bf t}', z^{-1})}\,
\hat \tau^{\rm TC}\Bigl (s+1, 
{\textstyle \frac{{\bf t}-[z^{-1}]}{2}}
\Bigr )\,
\hat \tau^{\rm TC}\Bigl (s', 
{\textstyle \frac{{\bf t}'+[z^{-1}]}{2}}
\Bigr )}
\\ \\
\phantom{aaaaaaaa}=\, \hat \tau^{\rm TC}(s, {\bf t}/2)\,
\hat \tau^{\rm TC}(s'+1, {\bf t'}/2).
\end{array}
\eeq
Comparing this 
with equation (\ref{L3}), it should be noted 
that in this case the second integral on the left side does not vanish, because there is a residue at infinity, and this term appears 
in the right-hand
side of (\ref{L10}).

Suppose first that $n\in 2\ZZ$ and $n'\in 2\ZZ +1$.
Then ansatz (\ref{ans}) allows us to rewrite equation (\ref{L1c})
in the form
\beq\label{L1d}
\begin{array}{l}
\displaystyle{
\phantom{aa}\oint_{C_{\infty}} \frac{dz}{z^2-1}\, 
z^{n-n'} e^{\xi ({\bf t}-{\bf t}', z)-\xi ({\bf t}-{\bf t}', z^{-1})}
\hat \tau^{\rm B} \Bigl ( n, {\bf t}-[z^{-1}]\Bigr )
\hat \tau^{\rm B} \Bigl ( n', {\bf t}'+[z^{-1}]\Bigr )}
\\ \\
\phantom{a}+\, \displaystyle{
\oint_{C_{\infty}} \frac{dz}{z^2-1}\, 
z^{n'-n} e^{-\xi ({\bf t}-{\bf t}', z)+\xi ({\bf t}-{\bf t}', z^{-1})}
\hat \tau^{\rm B} \Bigl ( n+2, {\bf t}+[z^{-1}]\Bigr )
\hat \tau^{\rm B} \Bigl ( n'-2, {\bf t}'-[z^{-1}]\Bigr )}
\\ \\
\phantom{aaaaaaaa}
=\, \tau^{\rm B}(n, {\bf t})\tau^{\rm B}(n', {\bf t}').
\end{array}
\eeq
Our task is to identify it with (\ref{L10}).
It is not difficult to see that it is enough to put
$$
s=\frac{n}{2}, \quad s'=\frac{n'-1}{2}
$$
and use (\ref{int1}).
The remaining case $n\in 2\ZZ +1$, $n'\in 2\ZZ$ 
is analyzed similarly.

\section{Examples}

In this section we give explicit examples of the correspondence
between the two hierarchies.
Using the example of multi-soliton solutions, we will clearly 
demonstrate how any $N$-soliton tau-function of the 1D TC becomes 
a large BKP tau-function.

It is instructive to start with the following well-known 
determinant formula for $N$-soliton solutions of the 2D TL:
\beq\label{e1}
\hat \tau^{\rm TL}(n, {\bf t}, \bar {\bf t})=
\det_{N\times N}\left (\delta_{ik} +
\frac{\alpha_i (p_i-q_i)}{p_i-q_k}\Bigl (p_i/q_i\Bigr )^n 
e^{\zeta ({\bf t}, \bar {\bf t}, p_i) -
\zeta ({\bf t}, \bar {\bf t}, q_i )} \right ),
\eeq
where
$$
\zeta ({\bf t}, \bar {\bf t}, p )=\xi ({\bf t}, p)-
\xi (\bar {\bf t}, p^{-1})
$$
and $p_i$, $q_i$, $\alpha_i$ ($i=1, \ldots , N$)
are arbitrary parameters (see e.g., \cite{UT84}).
It is also well known that the determinant 
can be found explicitly and this tau function can be 
represented as the following sum of $2^N$ terms:
\beq\label{e2}
\hat \tau^{\rm TL}(n, {\bf t}, \bar {\bf t})=
1+\sum_{m=1}^{N} \,\, \sum_{i_1< i_2 < \cdots < i_m}^N
\!\! \left ( \prod_{\gamma =1}^m \alpha_{i_{\gamma}}
\Bigl (\frac{p_{i_{\gamma}}}{q_{i_{\gamma}}}\Bigr )^n 
e^{\zeta ({\bf t}, \bar {\bf t}, p_{i_{\gamma}}) -
\zeta ({\bf t}, \bar {\bf t}, q_{i_{\gamma}})}\right )
\prod_{\mu <\nu}^m C_{i_\mu i_\nu} ,
\eeq
where
\beq\label{e3}
C_{ij}=\frac{(p_i-p_j)(q_j-q_i)}{(p_i-q_j)(p_j-q_i)}.
\eeq
For clarity, we will henceforth write down similar formulas, 
explicitly writing out only the first two terms in the sum over $m$:
\beq\label{e4}
\begin{array}{l}
\displaystyle{
\hat \tau^{\rm TL}(n, {\bf t}, \bar {\bf t})=
1+\sum_{i=1}^{N} \alpha_i \Bigl (\frac{p_i}{q_i}\Bigr )^n 
e^{\zeta ({\bf t}, \bar {\bf t}, p_{i}) -
\zeta ({\bf t}, \bar {\bf t}, q_{i})}}
\\ \\
\phantom{aaaaaaaaaaaa}\displaystyle{
+\,\, \sum_{i<j}^N \alpha_i \alpha_j C_{ij}
\Bigl (\frac{p_ip_j}{q_iq_j}\Bigr )^n 
e^{\zeta ({\bf t}, \bar {\bf t}, p_{i}) -
\zeta ({\bf t}, \bar {\bf t}, q_{i})
+\zeta ({\bf t}, \bar {\bf t}, p_{j}) -
\zeta ({\bf t}, \bar {\bf t}, q_{j})}\, + \ldots }
\end{array}
\eeq
The structure of other terms is clear from (\ref{e2}).
For example, at $m=3$ there is the contribution of the form
$$
\sum_{i<j<k}^N \alpha_i \alpha_j \alpha_k \, C_{ij}C_{ik}C_{jk}\, 
\Bigl (\frac{p_ip_jp_k}{q_iq_jq_k}\Bigr )^n\, 
e^{\zeta_i +\zeta_j +\zeta_k},
$$
where $\zeta_i =\zeta ({\bf t}, \bar {\bf t}, p_{i}) -
\zeta ({\bf t}, \bar {\bf t}, q_{i})$.

The reduction to the Toda chain corresponds to the choice
$q_i =p_i^{-1}$ for all $i$. In this case 
\beq\label{e5}
C_{ij}\, \longrightarrow \, 
C_{ij}^{\rm TC}=\left (\frac{p_i-p_j}{p_ip_j-1}
\right )^2,
\eeq
and
\beq\label{e6}
\zeta ({\bf t}, \bar {\bf t}, p) -
\zeta ({\bf t}, \bar {\bf t}, p^{-1})=2\xi ({\bf T}, p)-
2\xi ({\bf T}, p^{-1})\equiv 2 \zeta ({\bf T}, p), 
\qquad {\bf T}={\textstyle \frac{1}{2}}
({\bf t}-\bar {\bf t}).
\eeq
Therefore, for the tau-function $\hat \tau^{\rm TC}$ we can write:
\beq\label{e7}
\begin{array}{l}
\displaystyle{
\hat \tau^{\rm TC}(n/2, {\bf T}/2)=
1+\sum_{i=1}^{N} \alpha_i \, p_i^n\,
e^{\xi ({\bf T}, p_{i}) -
\xi ({\bf T}, p^{-1}_{i})}}
\\ \\
\phantom{aaaaaaaaaaaa}\displaystyle{
+\,\, \sum_{i<j}^N \alpha_i \alpha_j \, C^{\rm TC}_{ij}\, 
(p_ip_j)^n 
e^{\xi ({\bf T}, p_{i}) -
\xi ({\bf T}, p^{-1}_{i})
+ \xi ({\bf T}, p_{j}) -
\xi ({\bf T}, p^{-1}_{i})}\, + \ldots }
\end{array}
\eeq

Soliton solutions of the large BKP hierarchy were obtained
in \cite{PZ23}. They are expressed through Pfaffians rather 
than determinants. In the explicit form, tau-function of 
an $N$-soliton solution has a structure similar to (\ref{e2}),
(\ref{e4}) but the coefficients $C_{ij}$ are given by a different
formula. Namely, we have:
\beq\label{e8}
\begin{array}{l}
\displaystyle{
\hat \tau^{\rm B}(n, {\bf T})=
1+\sum_{i=1}^{N} \alpha_i \Bigl (p_i q_i\Bigr )^n 
e^{\zeta ({\bf T},  p_{i}) +
\zeta ({\bf T}, q_{i})}}
\\ \\
\phantom{aaaaaaaaaaaa}\displaystyle{
+\,\, \sum_{i<j}^N \alpha_i \alpha_j B_{ij}
\Bigl (p_ip_jq_iq_j\Bigr )^n 
e^{\zeta ({\bf T},  p_{i}) +
\zeta ({\bf T}, q_{i})+\zeta ({\bf T},  p_{j}) +
\zeta ({\bf T}, q_{j})
}\, + \ldots }\, ,
\end{array}
\eeq
where $\zeta ({\bf T}, p)=\xi ({\bf T}, p)-
\xi ({\bf T}, p^{-1})$ and
\beq\label{Bij}
B_{ij}=\frac{(p_i-p_j)\, (p_i-q_j)\, (q_i-p_j)\, (q_i-q_j)}{(p_ip_j-1)
(p_iq_j-1)(q_ip_j-1)(q_iq_j-1)}.
\eeq
We should compare (\ref{e7}) with (\ref{e8})
and try to match the $q_i$'s in this last expression 
so that it matches (\ref{e7}).
First of all we see that if $q_i\to 1$, then 
$\zeta ({\bf T}, q_i)\to 0$ and so the exponential factors
in (\ref{e7}) and (\ref{e8}) become the same in the limit. 
However, if $q_i$ and $q_j$ both 
tend to 1, there is an uncertainty in the coefficient $B_{ij}$
that needs to be resolved. To this end, we put
$$
q_i = 1 +\varepsilon \, \frac{p_i +1}{p_i-1}, \qquad \varepsilon \to 0,
$$
then
$$
\lim_{\epsilon \to 0} B_{ij} =
\left (\frac{p_i-p_j}{p_ip_j-1}
\right )^2= C_{ij}^{\rm TC},
$$
and so the tau-functions (\ref{e7}) and (\ref{e8})
become identical. We have explicitly demonstrated that 
any $N$-soliton tau-function of the TC is an
$N$-soliton tau-function of the large BKP. However, this
example shows that this is a rather special solution of the 
latter hierarchy, such that the parameters $q_i$ are fixed 
in a specific way.

At last, let us show how all this looks like
in the continuous limit in $n$, i.e., for the KdV and small
BKP hierarchies. The KdV tau-function for an $N$-soliton
solution is
\beq\label{e9}
\tau^{\rm KdV}({\bf t}_{o}/2)=
1+\sum_{i=1}^{N} \alpha_i 
e^{\xi_{o} ({\bf t}_{o}, p_{i})}
+\sum_{i<j}^N \alpha_i \alpha_j \, C^{\rm KdV}_{ij}\, 
e^{\xi_{o} ({\bf t}_{o}, p_{i})+\xi_{o} ({\bf t}_{o}, p_{j})}\, + 
\ldots  \, ,
\eeq
where
\beq\label{e10}
C^{\rm KdV}_{ij}=\left (\frac{p_i -p_j}{p_i+p_j}\right )^2.
\eeq
This formula is nothing else than (\ref{e4}) with the specialization
$n=0$ and $q_i=-p_i$ for all $i$.

Soliton solutions to the small BKP hierarchy were obtained
in \cite{Hirota89}. They are expressed through Pfaffians.
The structure of the formula is similar to (\ref{e8}).
The $N$-soliton tau-function reads
\beq\label{e11}
\tau^{\rm b}({\bf t}_{o})=
1+\sum_{i=1}^{N} \alpha_i 
e^{\xi_{o} ({\bf t}_{o}, p_{i})+\xi_{o} ({\bf t}_{o}, q_{i}) }
+\sum_{i<j}^N \alpha_i \alpha_j \, b_{ij}\, 
e^{\xi_{o} ({\bf t}_{o}, p_{i})+\xi_{o} ({\bf t}_{o}, q_{i})+
\xi_{o} ({\bf t}_{o}, p_{j})+\xi_{o} ({\bf t}_{o}, q_{j}) }\, + 
\ldots  \, ,
\eeq
where
\beq\label{e12}
b_{ij}=\frac{(p_i-p_j)(p_i-q_j)(q_i-p_j)(q_i-q_j)}{(p_i+p_j)
(p_i+q_j)(q_i+p_j)(q_i+q_j)}.
\eeq
This should be somehow identified with (\ref{e9}). This is indeed
possible if we set
$$
q_i =\varepsilon /p_i, \qquad \varepsilon \to 0,
$$
then $\xi_{o}({\bf t}_o , q_i)\to 0$ and
$$
\lim_{\varepsilon \to 0} b_{ij}=
\left (\frac{p_i -p_j}{p_i+p_j}\right )^2=
C^{\rm KdV}_{ij}.
$$
We see that with this choice, (\ref{e9}) and (\ref{e11}) match,
so any $N$-soliton KdV tau-function is an
$N$-soliton tau-function of the small BKP. Again, 
it is a rather special soliton solution of the 
latter hierarchy, such that the parameters $q_i$ tend to zero
in a specific way.

\section{Conclusion and outlook}

We have established a previously unknown relationship 
between the 1D Toda chain and large BKP regarded as 
infinite hierarchies of integrable 
differential-difference equations. Namely, we have proved that 
any solution of the former hierarchy is a solution of the
latter one (to be more precise, the simple rescaling 
${\bf t}\to {\bf t}/2$ of the time
variables is required for this). This relationship is
established on the level of tau-functions. The precise
statement is given by Theorem \ref{theorem:main1} 
(see equation (\ref{int1})).

In the both hierarchies, the space variable $n\in \ZZ$ 
is discrete, i.e., lives on the 1D lattice, 
and equations connecting fields at different points 
of the lattice are difference, not differential.
In this respect our result generalizes Alexandrov's work
\cite{A21}, where a similar relationship was established 
for the KdV and (small) BKP hierarchies, which can be 
regarded as continuous versions of the TC and large BKP.

It should be emphasized that solutions to large BKP 
coming from TC are rather special 
from the BKP point of view. (In Section 5 this point is illustrated
by the example of soliton solutions to the both hierarchies.) Indeed, 
the set of BKP solutions is much larger 
than the set of TC solutions: in the first case, the 
general solutions are parameterized by functions 
of two variables, while in
the second -- by functions of one variable.
A related special feature of large BKP tau-functions coming from
TC is that they are the same for the ``even''
($n\in 2\ZZ$) and ``odd'' ($n\in 2\ZZ +1$) sublattices. 
As it was already mentioned in Remark \ref{remark:dkp}, 
a copy of the DKP hierarchy is defined 
on each of the two sublattices. We thus see that any solution
that comes from TC yields a solution that is the same for 
the both copies.

If we limit ourselves to the soliton solutions discussed 
in Section 5, the result obtained seems rather surprising.
The fact is that, as is known, soliton solutions to BKP
(both large and small)
are expressed through Pfaffians, 
while solutions to KdV and Toda are 
expressed in terms of determinants.
Our result suggests that these determinants 
can equally well be represented as Pfaffians,
a rather unexpected fact that has apparently 
not been noted before.

In conclusion, we will note some questions for further research.
First of all, it is of interest to generalize the obtained result 
into multi-component and matrix versions of the hierarchies considered.
Second, it would be instructive to see how the 
correspondence found looks in the operator formalism 
when implementing the tau-functions as vacuum 
averages of certain fermionic operators.
Last but not least, it is worth studying examples of 
solutions originating from random matrix models 
in which tau-functions are their partition functions.
A related interesting point is to investigate how the
result of this note looks like in the dispersionless limit
of the hierarchies
(i.e., in the $N\to \infty$ limit of random matrix models).
We hope to address these problems elsewhere.

\section*{Acknowledgments}
\addcontentsline{toc}{section}{Acknowledgments}

I am grateful to A. Orlov, 
who always emphasized the importance of the 
large BKP hierarchy and its general and natural character.

This article is an output of the research project 
(HSE-BR-2025-84) implemented as a part of the 
Basic Research Program at the National Research University 
Higher School of Economics (HSE University).

\end{document}